\documentclass[prb,final,twocolumn,showpacs]{revtex4}
\usepackage{amssymb,amsmath}
\usepackage{bm}
\usepackage{dcolumn}
\usepackage{graphicx}
\begin{document}

\title{The influence of long-range correlated defects on critical ultrasound
propagation in solids}

\author{
\firstname{Pavel~V.}~\surname{Prudnikov},
\firstname{Vladimir~V.}~\surname{Prudnikov},
 }
\affiliation{%
Department of Theoretical Physics, Omsk State University,  Pr.Mira 55A, Omsk 644077, Russia
}%

\email{prudnikp@univer.omsk.su}

\date{\today}

\begin{abstract}
The effect of long-range correlated quenched structural defects on the critical ultrasound
attenuation and sound velocity dispersion is studied for three-dimensional Ising-like
systems. A field-theoretical description of the dynamic critical effects of
ultrasound propagation in solids is performed with allowance for both fluctuation
and relaxation attenuation mechanisms. The temperature and frequency dependences
of the dynamical scaling functions of the ultrasound critical characteristics are
calculated in a two-loop approximation for different values of the correlation parameter $a$
of the Weinrib-Halperin model with long-range correlated defects. The asymptotic behavior
of the dynamical scaling functions in hydrodynamic and critical regions is separated.
The influence of long-range correlated disorder on the asymptotic behavior of the critical
ultrasonic anomalies is discussed.
\end{abstract}


\maketitle

\section{Introduction}

The unique feature of ultrasonic methods consists in the fact that for a solid both
an anomalous peak of ultrasonic attenuation and an anomalous change in the
ultrasound velocity are observable from experiments
\cite{IkushimaF,Aliev} in vicinity of critical point.
These phenomena are caused by the interaction of low-frequency acoustic oscillations with
long-lived order parameter fluctuations, which produce a random force
that perturbs normal acoustic modes by means of magnetostrictive
spin-phonon interaction. In this process, relaxation and fluctuation
attenuation mechanisms can be distinguished. The relaxation
mechanism, which is due to a linear dynamic relationship between
sound waves and an order parameter,\cite{LandauKh} manifests itself
only in an ordered phase, where the statistical average of the order
parameter is nonzero. Since the relaxation of the order parameter
near a phase-transition point proceeds slowly, this mechanism plays
an important role in the dissipation of low-frequency acoustic
oscillations. The fluctuation attenuation mechanism, which is
determined by a quadratic relation between the deformation variables
in the Hamiltonian of a system with order parameter fluctuations,
manifests itself over the entire critical temperature range. To
date, there are a considerable number of works that deal with a
theoretical description of the ultrasonic critical anomalies,\cite{Pawlak,Schwabl93,Kamilov98} which give an adequate explanation
of experimental results.\cite{Bhatt,Luthi,Suzuki82}

One of the most interesting and important problems from both
experimental and theoretical viewpoints is the study of the
influence of structural defects on the dynamical process of sound propagation
in solids undergoing phase transitions. The structural disorder induced by impurities
or other structural defects plays a key role in the behavior of real materials and
physical systems. The structural disorder breaks the translational symmetry of
the crystal and thus greatly complicates the theoretical description of the material.
The influence of disorder is particularly important near critical point
where behavior of a system is characterized by anomalous large response
on any even weak perturbation. In most investigations consideration has
been restricted to the case of point-like uncorrelated defects.\cite{Folk_Hol}
However, the non-idealities of structure cannot be modeled by simple
uncorrelated defects only. Solids often contain defects of a more complex
structure: linear dislocations, planar grain boundaries, clusters of
point-like defects, and so on.

According to the Harris criterion,\cite{Harris74} the critical
behavior of only Ising systems is changed by the presence of
quenched point-like defects. The problem of the influence of
point-like defects on the critical sound propagation in Ising
systems has been discussed in paper \cite{PawlakFecher89} with the use
of a $\varepsilon$--expansion in the lowest order of approximation.
However, our pilot analysis of this phenomenon showed that in
Ref.~\onlinecite{PawlakFecher89} some diagrams which are needed for a
correct description of the influence of the disorder were not
considered. Furthermore, our numerous investigations of pure and
disordered systems performed in the two-loop and higher orders of
the approximation for the three-dimensional (3D) system directly,
together with the use of methods of series summation, show that the
predictions made in the lowest order of the approximation,
especially on the basis of the $\varepsilon$ expansion, can differ
strongly from the real critical behavior.\cite{Prudnikov_JETPL97,Prudnikov_JETP98,Prudnikov_JETP99,PrudnikovPR00,PrudnikovPR01,Prudnikov_JETP02}
In paper \cite{PrudnikovCM} we have realized a correct
field-theoretical description of dynamical effects of the influence of
point-like defects on acoustic anomalies near temperature of the
second-order phase transition for 3D Ising systems in
the two-loop approximation with consideration of only the
fluctuation attenuation mechanism. In papers \cite{Prudnikov_PhMM,Prudnikov_JETP} we
have performed, the field-theoretical description of the influence of
point-like defects on critical acoustic anomalies with allowance for
both fluctuation and relaxation attenuation mechanisms.

The temperature and frequency dependencies of the scaling functions for
ultrasonic attenuation and sound velocity dispersion were calculated
in paper\cite{Prudnikov_JETP} the two-loop approximation with the use of the resummation of asymptotic
series technique for pure and disordered systems, and
their asymptotic behavior in hydrodynamic and critical regions were
distinguished. It was shown that the presence of structural defects
causes a stronger temperature and frequency dependencies of the
acoustic characteristics in the critical region and increase in the
critical anomalous peak of ultrasonic attenuation in comparison with
pure system.

However, the question about influence of long-range (LR) correlation
effects realizing in defects with complex structure on critical
behavior of systems was investigated noticeably weaker, and this
question has not been studied entirely in respect to display of
correlation effects in the critical anomalous properties of sound
propagation in solids. Different models of structural disorder have
arisen as an attempt to describe complicated defects. In this paper,
we concentrate on the model of Weinrib and Halperin (WH) \cite{WH}
with the so-called LR-correlated disorder when pair
correlation function for point-like defects $g({\bm{x}}-{\bm{y}})$
falls off with distance as a power law $g({\bm{x}}-{\bm{y}})\sim
|{\bm{x - y}}|^{-a}$. Weinrib and Halperin showed that for $a\geq d$
LR correlations are irrelevant and the usual short-range (SR)
Harris criterion \cite{Harris74} $2 - d\nu_0 = \alpha_0 > 0$ of the
effect of point-like uncorrelated defects is realized, where $d$ is
the spatial dimension, and $\nu_0$ and $\alpha_0$ are the
correlation-length and the specific-heat exponents of the pure
system. For $a < d$ the extended criterion $2 - a\nu_0 > 0$ of the
effect of disorder on the critical behavior was established. As a
result, a wider class of disordered systems, not only the
3D diluted Ising model with point-like uncorrelated
defects, can be characterized by a different type of critical behavior.
So, for $a < d$  a LR-disorder stable fixed point
(FP) of the renormalization group (RG) recursion relations for systems
with a number of components of the order parameter $m \geq 2$ was
discovered. The critical exponents were calculated in the one-loop
approximation using a double expansion in $\varepsilon = 4 - d \ll
1$ and $\delta = 4 - a \ll 1$. The correlation-length exponent was
evaluated in this linear approximation as $\nu=2/a$ and it was
argued that this scaling relation is exact and also holds in higher
order approximation. In the case $m = 1$ the accidental degeneracy
of the recursion relations in the one-loop approximation did not
permit to find LR-disorder stable FP. Korzhenevskii {\it et al}.
\cite{Korzhenevskii} proved the existence of the LR-disorder stable
FP for the one-component WH model and also found characteristics of
this type of critical behavior.

The results for WH model with LR correlated defects received based
on the $\varepsilon, \delta$ expansion
\cite{WH,Korzhenevskii,Dorogovtsev,Uzunov,Korucheva} were questioned
in our paper,\cite{PrudnikovPR00} where a renormalization analysis
of scaling functions was carried out directly for the 3D systems in
the two-loop approximation with the values of $a$ in the range
$2\leq a \leq 3$, and the FP's corresponding to stability of various
types of critical behavior were identified. The static and dynamic
critical exponents in the two-loop approximation were calculated
with the use of the Pad\'{e}-Borel summation technique. The results
obtained in Ref.~\onlinecite{PrudnikovPR00} essentially differ from the results
evaluated by a double $\varepsilon, \delta$ - expansion. The
comparison of calculated the exponent $\nu$ values and ratio $2/a$
showed the violation of the relation $\nu = 2/a$, assumed in
Ref.~\onlinecite{WH} as exact. Monte-Carlo simulations of the short-time
dynamic behavior were carried out in paper \cite{Prudnikov_PTP} for
3D Ising and \textit{XY} models with LR correlated disorder at criticality,
in the case corresponding to linear defects with $a=2$. It was shown
that the obtained values of the static and dynamic critical exponents
are in a good agreement with our results of the field-theoretic description
of the critical behavior of these models in Ref.~\onlinecite{PrudnikovPR00}.

The models with LR-correlated quenched defects have both theoretical
interest due to the possibility of predicting types of critical
behavior in disordered systems and experimental interest due to the
possibility of realizing LR-correlated defects in the orientational
glasses,\cite{Binder} polymers,\cite{Blavats'ka} and disordered
solids containing fractal-like defects,\cite{Korzhenevskii}
dislocations near the sample surface \cite{Altarelli} or $\mathop{\mathrm{{}^4He}}$
in porous media such as silica aerogels and xerogels.\cite{He4}
 For this purpose, in this work we performed a field-theoretical description of
the effect of LR-correlated quenched defects on the anomalous
critical ultrasound attenuation in 3D systems with allowance for both
the fluctuation and relaxation attenuation mechanisms without using
the $\varepsilon, \delta$ expansion. We considered in this paper only
Ising-like systems that gives to compare the effect of LR-correlated defects
with point-like uncorrelated defects on the critical ultrasound attenuation.

In Sec.~II, we describe the model of compressible Ising-like systems with
LR-correlated disorder and set of dynamical equations which are used for
description of the critical sound propagation in solids. We also give
in Sec.~II the results of RG expansions in a two-loop approximation
for ultrasound scaling functions, and $\beta$ and $\gamma$ functions of
the Callan-Symanzik RG equations. The FP's corresponding to the stability of
various types of critical behavior and the critical exponents are determined
in Sec. III. Section IV is devoted to analysis of the ultrasound critical
characteristics and discussion of the main results.

\section{Model and description of dynamical equations. RG expansions}

For description of the critical sound propagation in solids the
model of compressible media is used. Interaction of the order
parameter with elastic deformations plays a significant role in the
critical behavior of the compressible system. It was shown for the
first time in Ref.~\onlinecite{Larkin69} that the critical behavior of a
system with elastic degrees of freedom is unstable with respect to
the connection of the order parameter with acoustic modes and a
first-order phase transition is realized. However, the conclusions
of Ref.~\onlinecite{Larkin69} are only valid at low pressures. It was shown
in Ref.~\onlinecite{Ymry74} that in the range of high pressures, beginning
from a threshold value of pressure, the deformational effects
induced by the external pressure lead to a change in type of the
phase transition.

The Hamiltonian of a disordered compressible Ising model can be
written as
\begin{equation} \label{ham:1}
H = H_{\mathrm{el}} + H_{\mathrm{op}} + H_{\mathrm{int}} + H_{\mathrm{imp}},
\end{equation}
consisting of four contributions.

The contribution $H_{\mathrm{el}}$ in (\ref{ham:1}) of the deformation degrees of
freedom is determined as
\begin{multline}
H_{el}=\frac{1}{2}\int {\rm d^{d}}x\, \left(C^{0}_{11}\sum\limits_{\alpha} u^{2}_{\alpha \alpha} \right. +\\
  +\,2C^{0}_{12}\sum\limits_{\alpha\beta} u_{\alpha \alpha} u_{\beta \beta}
  + \left. 4C^{0}_{44} \sum\limits_{\alpha <\beta}u^{2}_{\alpha\beta}\right),
\end{multline}
where $u_{\alpha \beta}(x)$ are components of the strain tensor and
$C_{ij}^{k}$ are the elastic moduli. The use of an isotropy
approximation for $H_{el}$ is caused by the fact that in the
critical region the behavior of Hamiltonian parameters of anisotropic system is determined
by an isotropic FP of RG transformations, while the anisotropy effects are negligible.\cite{Izym}
This phenomenon of the increase in system symmetry in the critical point is well known as
asymptotic symmetry.

$H_{\mathrm{op}}$ is a magnetic part in the appropriate Ginzburg–Landau-Wilson form:
\begin{equation}
H_{op} = \int  {\rm d^{d}}x\, \left[ \frac{1}{2}\tau_0 S^{2} + \frac{1}{2}\left(\nabla S\right)^{2} +
\frac{1}{4}u_0 S^{4} \right],
\end{equation}
where $S(x)$ is the Ising-field variable which is associated with the spin order parameter,
$u_0$ is a positive constant and $\tau_0 \sim (T - T_{0c})\left/ T_{0c}\right.$ with the
mean-field phase transition temperature $T_{0c}$.

The term $H_{\mathrm{int}}$ describes the spin-elastic interaction
\begin{equation}
H_{int} = \int {\rm d^{d}}x\, \left[ g_{0} \sum_{ \alpha} u_{\alpha \alpha} S^{2} \right],
\end{equation}
which is bilinear in the spin order parameter and linear in deformations.
$g_{0}$ is the bare coupling constant.

The term $H_{\mathrm{imp}}$ of the Hamiltonian determines the influence of disorder and
it is considered in the following form:
\begin{equation}
H_{imp} = \int {\rm d^{d}}x\, \left[\Delta{\tau}(x)S^{2}\right]+\int{\rm d^{d}}x\, \left[h(x)\sum\limits_{\alpha} u_{\alpha\alpha}\right],
\end{equation}
where the random Gaussian variables $\Delta{\tau}(x)$ and $h(x)$ determine the local
transition temperature fluctuations and induced random stress, respectively.
The second moment of distribution of the $\Delta{\tau}(x)$ fluctuations
$g({\bm{x}}-{\bm{y}})=\langle\langle\Delta{\tau}(\bm{x})\Delta{\tau}(\bm{y})\rangle\rangle \sim |{\bm{x - y}}|^{-a}$
characterizes the LR correlation effects realizing in defects with complex structure.
The second moment of distribution of the $h(x)$ fluctuations
$C({\bm{x}}-{\bm{y}})=\langle\langle h(\bm{x})h(\bm{y})\rangle\rangle$ presets a
renormalization of the elastic moduli and the coupling constant in the spin-elastic
interaction.

To perform calculations, it is convenient to use the
Fourier components of the deformation variables in the
form:
\begin{equation}
\label{eq:Ftr}
 u_{\alpha\beta} = u^{(0)}_{\alpha \beta} + V^{-1/2} \sum_{q \neq 0}
 u_{\alpha\beta}(q) \exp\left(i q x\right),
\end{equation}
where $q$ is the wavevector, $V$ is the volume, $u^{(0)}_{\alpha \beta}$ is an
uniform part of the deformation tensor, and $u_{\alpha \beta}(q) = {\rm i}/2\left[q_\alpha u_\beta + q_\beta u_\alpha\right]$.
Then the normal-mode expansion is introduced as
\begin{equation*}
\mathbf{u}(q)=\sum_\lambda \mathbf{e}_\lambda(q) Q_{q,\lambda},
\end{equation*}
with the normal coordinate $Q_{q,\lambda}$ and polarization vector $\mathbf{e}_\lambda(q)$.
We carry out the integration in the partition function with respect to the nondiagonal
components of the uniform part of the deformation
tensor $u^{(0)}_{\alpha \beta}$, which are insignificant for the critical behavior
of the system in an elastically isotropic medium.
After all of the transformations, the effective Hamiltonian of the system
has become in the form of a functional for the spin
order parameter $S\left( q\right)$ and the normal coordinates $Q_\lambda\left( q\right)$:
\begin{gather}
\label{ham:8}
\tilde{H} = \displaystyle\frac{1}{2}\int {\rm d^{d}}q\ \left(\tau_0+q^{2}\right)\, S_{q}\, S_{-q}
          + \displaystyle\int {\rm d^{d}}q\ q\,h_{q}\, Q_{-q,\lambda}  + \nonumber \\
      + a_0\displaystyle\int {\rm d^{d}}q\ q^{2}\, Q_{q,\lambda}\, Q_{-q,\lambda}
          + \displaystyle\frac{1}{2}\int {\rm d^{d}}q\ \Delta{\tau}_{q}\, S_{q_1}\, S_{q_1-q} + \nonumber \\
      + \displaystyle\frac{1}{4}u_0 \int {\rm d^{d}}q\ S_{q_1}\, S_{q_2}\, S_{q_3}\, S_{-q_1-q_2-q_3}- \\
       - \displaystyle \frac{1}{2}\mu_0 \int {\rm d^{d}}q\ \left(S_{q}\, S_{-q}\right)\, \left(S_{q}\, S_{-q}\right)- \nonumber\\
          -g_0\displaystyle\int {\rm d^{d}}q\ q\, Q_{-q,\lambda}\, S_{q_1}\, S_{q-q_1}.  \nonumber
\end{gather}
where
\begin{align}
\mu_0&=\displaystyle\frac{3g_0^2}{V\left( 4C_{12}^0 - C_{11}^0 \right)},
&a_0&=\displaystyle\frac{C_{11}^0+4C_{12}^0-4C_{44}}{4V}. \nonumber
\end{align}
The Fourier transformation of the correlation function for distribution defects in system
$g(\bf x) \sim |\bf x|^{-a}$ gives $g({\bf k})= v_0 + w_0 k^{a-d}$ for small $k$. $g(\bf k)$
must be positive definite, therefore, if $a>d$, then the $w_0$ term is irrelevant,
$v_0 \geq 0$, and the Hamiltonian (7) describes the model with SR-correlated
defects, while if $a<d$, then the $w_0$ term is dominant at small $k$ and
$w_0 \geq 0$ and characterizes a large influence of LR-correlated defects on the
critical behavior.

The critical dynamics of the compressible
system in the relaxational regime can be described by the Langevin equations \cite{MSI}
for the spin order parameter $S_{q}$ and phonon normal coordinates $Q_{q,\lambda}$:
\begin{eqnarray}
\label{dynamic:1}
& &\dot{S}_{q}=-\Gamma_{0}\,\frac{\partial\tilde{H}}{\partial S_{-q}}+\xi_{q}+\Gamma_{0} h_{S}, \\
\label{dinamic:2}
& &\ddot{Q}_{q,\lambda}=-\frac{\partial\tilde{H}}{\partial Q_{-q,\lambda}} -
q^{2} D_{0} \dot{Q}_{q,\lambda}+\eta_{q}+h_{Q}, \nonumber
\end{eqnarray}
where $\Gamma_{0}$ and $D_{0}$ are the bare kinetic coefficients,
$\xi_{q}(x,t)$ and $\eta_{q}(x,t)$ are Gaussian white noises, and
$h_S$ and $h_Q$ are the fields thermodynamically conjugated to the spin
and deformation variables, respectively.

The quantities of interest are the response functions $G(q,\omega)$ and $D(q,\omega)$
of the spin and deformation variables, respectively. It can be obtained by linearization
in correspondent fields that
\begin{eqnarray}
D(q,\omega) &=& \displaystyle\frac{\delta \left[\langle{Q_{q,\omega,\lambda}}\rangle\right]}{\delta{h_{Q}}}
= \left[\langle Q_{q,\omega,\lambda}Q_{-q,-\omega,\lambda}\rangle\right], \medskip \\
G(q,\omega) &=& \displaystyle\frac{\delta \left[\langle{S_{q,\omega}}\rangle\right]}{\delta{h_{S}}}
= \left[\langle S_{q,\omega}S_{-q,-\omega}\rangle\right],
\end{eqnarray}
where $\langle ... \rangle$ denotes averaging over Gaussian white noises,
$\left[ ... \right]$ denotes averaging over random fields
$\Delta{\tau}_{-q}$ and $h_{q}$ that are specified
by structural defects.

The response functions may be expressed in terms of self-energy parts:
\begin{eqnarray}
\label{Dys1}
G^{-1}(q,\omega) &=& G_{0}^{-1}(q,\omega) + \Pi(q,\omega), \\
D^{-1}(q,\omega) &=& D_{0}^{-1}(q,\omega) + \Sigma(q,\omega), \nonumber
\end{eqnarray}
where the free response functions $G_0(q,\omega)$ and $D_0(q,\omega)$ have the forms
\begin{eqnarray}
D_{0}(q,\omega,\lambda) &=& 1\left/\left(\omega^{2}-a q^{2}-i\omega D_{0} q^{2}\right)\right., \nonumber\\
G_{0}(q,\omega) &=& 1\left/\left[i\omega\left/\Gamma_{0}\right.+\left(\tau_0+q^{2}\right)\right]\right.. \nonumber
\end{eqnarray}

In the low-temperature phase, the response functions contain
an additional relaxation contribution caused by the presence in the spin density
\begin{equation}
\label{dinamic:1f}
 S_{q}=M\delta_{q,0}+\varphi_{q}
\end{equation}
of an additional part connected with the magnetization
\begin{equation}
\label{dinamic:2f}
 M = \left\{
\begin{array}{ll}
0, &  T > T_{c}, \\
B\left|T-T_{c}\right|^{\beta}, &  T < T_{c},
\end{array}
\right.
\end{equation}
where $B$ is the phenomenological relaxation parameter
and $\varphi_{q}$ is the fluctuation part of the order parameter.

\begin{figure*}
\includegraphics[width=\textwidth]{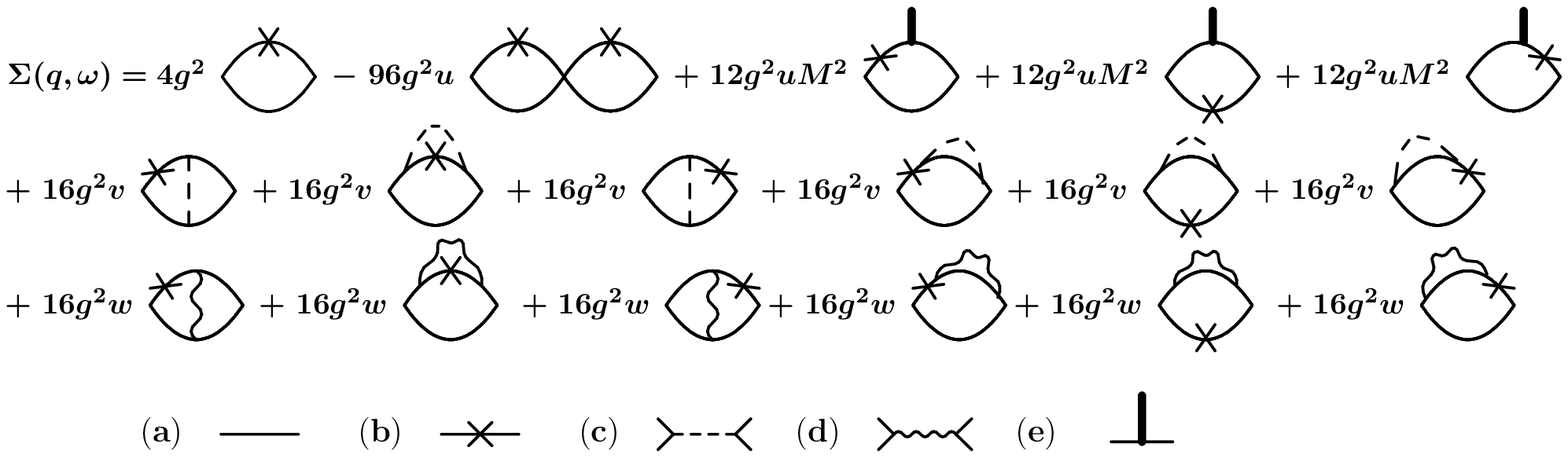}
\caption{ \label{fig:1} Diagrammatic representation of
$\Sigma(q,\omega)$ in a two-loop approximation.
($\mathbf{a}$) Solid line corresponds to $G_{0}(q,\omega)$;
($\mathbf{b}$) lines with a cross, to
$C_{0}(q,\omega)=2\Gamma_{0}^{-1}\left[
\left(\omega/\Gamma_0\right)^2+\left(q^2+\tau_0\right)^2\right]^{-1}$;
($\mathbf{c}$) vertex with dashed line, to
$v=\bigl[(\Delta \tau)^2\bigr]$;
($\mathbf{d}$) vertex with waved line, to
$w$, which characterized long-range correlation of disorder with wave vector $q^{a-d}$;
and ($\mathbf{e}$) line, to
relaxation insertion $M^2\delta_{q,0}$. }
\end{figure*}

The characteristics of the critical sound propagation are defined by means of the
response function \cite{IroSchwabl} $D(q,\omega)$.
Thus, the coefficient of ultrasonic attenuation is determined through
the imaginary part of $\Sigma(q,\omega)$:
\begin{equation}
\label{Atten}
    \alpha(\omega,\tau) \sim \omega\,{\rm Im}\Sigma(0,\omega),
\end{equation}
and the sound velocity dispersion is expressed through
its real part,
\begin{equation}
\label{disp}
    c^{2}(\omega,\tau) - c^{2}(0,\tau) \sim \mbox{Re}\left[\Sigma(0,\omega)-\Sigma(0,0)\right].
\end{equation}
We calculated $\Sigma(q,\omega)$ in the two-loop approximation. The diagrammatic
representation of $\Sigma(q,\omega)$ is presented in Fig.~\ref{fig:1}.

The Feynman diagrams involve momentum integrations in dimension $d$
(in our case, $d = 3$). Near the critical point the correlation length $\xi$ increases
infinitely. When $\xi^{-1} \ll \Lambda$, where $\Lambda$ is a cut-off in
the momentum-space integrals (the cut-off $\Lambda$ serves to specify the basic
length scale), the vertex functions are expected to display an asymptotic
scaling behavior for wavenumbers $q \ll \Lambda$. Therefore, one is lead to
consider the vertex functions in the limit $\Lambda \rightarrow \infty$.
The use of the RG scheme removes all divergences which arise
in thermodynamic variables and kinetic coefficients in this limit.

To calculate the attenuation coefficient (\ref{Atten}) and the ultrasound
velocity dispersion (\ref{disp}) and to eliminate the divergences in
 $\Sigma(q,\omega)$ at $q\to~0$, we used the matching
method which was introduced for statics in Ref.~\onlinecite{Nelson76} and then generalized
for critical dynamics in Ref.~\onlinecite{Matching}.
First, we use the dynamical scaling property of the response function
\begin{equation}
\label{RespF}
D\left(q,\omega,\tau\right) = e^{\left(2-\eta\right) l}D\left(q e^{l},\left(\omega/\Gamma_0\right) e^{zl},\tau e^{l/\nu}\right),
\end{equation}
and then we calculate the right-hand side of this equation for some value $l^*=l$,
where the arguments do not all vanish simultaneously.\cite{Nelson76}
The choice of $l^*$ is determined by the condition
\begin{equation}
\label{MCond}
\left[ \left( \omega/\Gamma_0 \right) e^{zl^*} \right]^{4/z} +
\left[ \left( \tau e^{l^*/\nu} \right)^{2\nu} + q^2 e^{2l^*}\right]^2=1,
\end{equation}
which permits to achieve the main purpose of the RG transformations,
namely, to find a relation between the behavior of the system in the precritical regime
at a low value of the reduced temperature $\tau$ and the behavior of the system in a
regime far from the critical mode, i.e., without divergences in
$\Sigma(q,\omega)$. As it was demonstrated in Ref.~\onlinecite{Matching}, the matching
condition (\ref{MCond}) provides an infrared cut-off for all
diverging quantities. The particular form of the matching condition (\ref{MCond})
permits an explicit solution for $l^*$ in the form of a functional dependence on
$\omega$ and $\tau$, which is specified by the dynamic critical
exponent $z$ and the static critical exponent $\nu$ for the correlation length,
\begin{equation}
\label{el}
e^{l^*}=\tau^{-\nu} \left[ \,1 +\left(y/2\right)^{4/z} \right]^{-1/4}\equiv \tau^{-\nu}F(y),
\end{equation}
where the abbreviation $y=\omega\tau^{-z \nu}\left/\Gamma_{0}\right.$ is introduced as
the argument of the $F(y)$ function.

The response function $D\left[q e^{l},\left(\omega/\Gamma_0\right) e^{zl},\tau e^{l/\nu}\right]$
on the right-hand side of (\ref{RespF}) is represented by Dyson Eq.~(\ref{Dys1})
and the following scaling relationships for the imaginary and real components of
self-energy part are valid:\cite{IroSchwabl}
\begin{eqnarray} \label{ImSelfR}
& \displaystyle\frac{ {\rm Im}\Sigma(\omega)}{\omega} =
e^{l {\left(\alpha+z\nu\right)}\left/{\nu}\right.}
\displaystyle\frac{ {\rm Im}\Sigma(\omega e^{zl})}{\omega e^{zl}},
\end{eqnarray}
\begin{multline}
\label{ReSelfR}
\mathrm{Re}\left[\Sigma(0,\omega)-\Sigma(0,0)\right]=\\
=e^{l \alpha\left/{\nu}\right.}\mathrm{Re}\left[\Sigma(0,\omega e^{zl})-\Sigma(0,0)\right],
\end{multline}
where $\alpha$ is the exponent for heat capacity.
It may be argued \cite{Matching} that condition (\ref{MCond}) with the well-known
expression for the susceptibility \cite{PawlakFecher89} provides an infrared cut-off
for all divergent values.

It was shown in later theoretical works \cite{Kawasaki,Suzuki82} that in asymptotic regions
with $\tau \to 0$ and $\omega \to 0$ the coefficient of attenuation and the sound velocity
dispersion are described by a simple scaling functions of the variable
$y$ only. The experimental investigations performed on the crystals Gd \cite{Luthi}
and MnP \cite{Suzuki82} confirmed the validity of the concepts of dynamical scaling.

In the asymptotic limit ($\tau \to 0$,\ $\omega \to 0$),
the expression for the imaginary part of $\Sigma(\omega,\tau)$ can be defined
by a scaling function $\phi(y)$
\begin{equation}
\label{ImScF}
\mbox{Im}\Sigma(\omega)\left/\omega\right. = \tau^{-\alpha-z\nu}\phi(y),
\end{equation}
and the expression for the real part of $\Sigma(\omega,\tau)$ can be
determined using another scaling function $f(y)$
\begin{equation}
\label{ReScF}
\mathrm{Re}\left[\Sigma(0,\omega)-\Sigma(0,0)\right]=\tau^{-\alpha}\left[f(y)-f(0)\right].
\end{equation}
The substitution of $e^{l^*}$ from Eq.~(\ref{el}) into the
right-hand side of the expressions (\ref{ImSelfR}) and (\ref{ReSelfR}) allows
to calculate the $\phi(y)$ and $f(y)$ scaling functions.

The dynamic scaling functions calculated in the two-loop
approximation have the forms:
\begin{eqnarray}
\label{ScaleFphi}
\phi(y) &=&
\displaystyle\frac{g^{*2}\Gamma_0}{\pi\mathstrut}
    \displaystyle\frac{F^{\alpha/\nu+1/2\nu-z}}{y^2}
\left[ 1-  \displaystyle\frac{\left(\Delta+1\right)^{1/2}}{\sqrt{2}}  \right] \nonumber \\
 &-&\displaystyle\frac{ 3g^{*2}u^*\Gamma_0^2}{\sqrt{2}\pi^2\mathstrut}
    \displaystyle\frac{ F^{\alpha/\nu+1/\nu-2 z}}{ y^3}
        \left(\Delta -1 \right)^{1/2}\\
&\times& \left[1- \displaystyle\frac{\left(\Delta+1\right)^{1/2}}{\sqrt{2}} \right]-\displaystyle\frac{g^{*2}v^*\Gamma_0}{12\,\pi^3}
        \displaystyle\frac{F^{\alpha/\nu-z}}{y^2}
        \,\ln\,\Delta
 \nonumber \\
 &-& M^{2}
\frac{3g^{* 2}u^*\Gamma_0}{2\pi}
\frac{F^{\alpha/\nu-1/2\nu-z}}{y^{2}}
\left[1-\frac{\left(\Delta+1\right)^{1/2}}{\sqrt{2}\Delta}\right]\nonumber\\
&-&\frac{w^*\Gamma_{0}^{2}}{4\,\pi^3}\frac{ F^{(\alpha+a+1)/\nu-z}}{ y^2}
 \frac{\Gamma(a-1)\Gamma(3/2-a/2)}{\Gamma(3/2)}\nonumber\\
&\times&
\Gamma(1-a/2)\left\{1-\cos\left[\frac{a-3}{2}\arctan\left(\frac{y\,F^{1/\nu-z}}{2}\right)\right]\right\}, \nonumber
\end{eqnarray}
\begin{eqnarray}
\label{ScaleFf}
 f(y) &=&
 \displaystyle\frac{g^{*2}\Gamma_0^2}{\pi\mathstrut}
\displaystyle\frac{F^{\alpha/\nu+1/2\nu-z}}{y}
\left[  \displaystyle\frac{\left(\Delta-1\right)^{1/2}}{\sqrt{2}} \right] \nonumber \\
 &-&\displaystyle\frac{ 3\,g^{*2}u^*\Gamma_0^3}{\pi^2\mathstrut}
    \displaystyle\frac{ F^{\alpha/\nu+1/\nu-2 z}}{ y^2}
        \left[  \displaystyle\frac{\left(\Delta+1\right)^{1/2}}{\sqrt{2}} -1 \right] \\
&-& M^2\frac{3g^{*2}u^*\Gamma_0^2}{2\pi}\frac{F^{\alpha/\nu-1/2\nu-z}}{ y}
\left[\frac{\left(\Delta-1\right)^{1/2}}{\Delta\sqrt{2}}\right]\nonumber\\
 &+& \displaystyle\frac{g^{*2}v^*\Gamma_0^2}{12\pi^3}
        \displaystyle\frac{F^{\alpha/\nu-z}}{y}
        \,\arctan(\Delta^{2}-1)^{1/2}
 \nonumber \\
&-&\frac{w^*\Gamma_{0}^{2}}{4\,\pi^3}\frac{ F^{(\alpha+a+1)/\nu-z}}{ y^2}
 \frac{\Gamma(a-1)\Gamma(3/2-a/2)}{\Gamma(3/2)}\nonumber\\
& \times &
\Gamma(1-a/2)\sin\left[\,\frac{a-3}{2}\arctan\left(\frac{y\,F^{1/\nu-z}}{2}\right)\right], \nonumber
\end{eqnarray}
\begin{eqnarray}
 \Delta &=& \left[1+\displaystyle\frac{y^2F^{2z-2/\nu}}{4}\right]^{1/2}, \nonumber
\end{eqnarray}
where $g^*$, $u^*$, $v^*$ and $w^*$ are  values of the interaction vertices in
the FP of the RG transformations that corresponds to the critical behavior of the
disordered compressible Ising model,\cite{Prudnikov01} generalized in this paper
to the case with LR-correlated disorder.
The terms in Eqs.~(\ref{ScaleFphi}) and (\ref{ScaleFf}) that are proportional to
$M^2$ describe the relaxation contribution for the scaling functions
of the attenuation coefficient and the sound velocity dispersion.

\begin{table*}
\caption{Coefficients for the $\beta$-functions in Eq.~\protect{\ref{eq:beta}}.}
\label{tab:b1}
\begin{tabular}{dddddddd}\hline\hline
\multicolumn{1}{c}{$a$}&\multicolumn{1}{c}{$b_1$}&\multicolumn{1}{c}{$b_2$}&\multicolumn{1}{c}{$b_3$}
&\multicolumn{1}{c}{$b_4$}&\multicolumn{1}{c}{$b_5$}&\multicolumn{1}{c}{$b_6$} &\multicolumn{1}{c}{$b_7$}   \\ \hline
3.00  &   1.851\,852  &  9.703\,704  &  3.425\,924 &   6.851\,852 &  1.000\,000 & 2.000\,000 & 2.666\,664   \\
2.90  &   1.751\,381  &  9.149\,428  &  3.047\,432 &   6.456\,936 &  1.632\,512 & 1.836\,920 & 2.376\,632   \\
2.80  &   1.662\,830  &  8.671\,819  &  2.747\,992 &   6.117\,964 &  2.148\,800 & 1.696\,968 & 2.190\,416   \\
2.70  &   1.584\,520  &  8.260\,292  &  2.508\,396 &   5.827\,228 &  2.584\,256 & 1.576\,244 & 2.086\,520   \\
2.60  &   1.515\,077  &  7.906\,550  &  2.313\,792 &   5.578\,740 &  2.967\,744 & 1.471\,748 & 2.053\,896   \\
2.50  &   1.453\,357  &  7.604\,029  &  2.151\,672 &   5.367\,788 &  3.325\,888 & 1.381\,120 & 2.089\,720   \\
2.40  &   1.398\,383  &  7.347\,527  &  2.010\,060 &   5.190\,684 &  4.022\,776 & 1.302\,500 & 2.199\,488   \\
2.30  &   1.349\,314  &  7.132\,943  &  1.875\,288 &   5.044\,576 &  4.904\,458 & 1.234\,384 & 2.396\,896   \\
2.20  &   1.305\,402  &  6.957\,111  &  1.728\,540 &   4.927\,324 &  5.918\,144 & 1.175\,580 & 2.708\,976   \\
2.10  &   1.265\,968  &  6.817\,670  &  1.539\,252 &   4.837\,412 &  7.035\,808 & 1.125\,124 & 3.183\,336   \\
2.00  &   1.230\,378  &  6.713\,001  &  1.250\,616 &   4.773\,916 &  8.216\,616 & 1.082\,260 & 3.905\,832   \\  \hline\hline
\end{tabular}
\end{table*}

In accordance with our paper,\cite{Prudnikov01} the values of
$g^*$, $u^*$, $v^*$, and $w^*$ can be obtained from the RG considerations of
the replicated Hamiltonian for the disordered compressible Ising model in which
we averaged over deformation variables. The final form of this Hamiltonian
can be presented as
\begin{gather}
\label{Hr}
H_R = \displaystyle\frac{1}{2}\int {\rm d^{d}}q\ \left(\tau_0+q^{2}\right)\, \sum_{a=1}^n S_{q}^a\, S_{-q}^a \nonumber + \\
      + \displaystyle\frac{u_0 - 2g_0^2}{4} \sum_{a=1}^n \int {\rm d^{d}}\{q_i\} S_{q_1}^a\, S_{q_2}^a\, S_{q_3}^a\, S_{-q_1-q_2-q_3}^a + \nonumber\\
       + \displaystyle \frac{g_0^2 - \tilde{\mu}_0}{2} \sum_{a=1}^n \int {\rm d^{d}}\{q_i\} \left(S_{q_1}^a\, S_{-q_1}^a\right)\, \left(S_{q_2}^a\, S_{-q_2}^a\right) - \label{eq:HR}\\
        - \displaystyle \frac{1}{2} \sum_{a,b=1}^n \int {\rm d^{d}}\{q_i\} g(q_1+q_2) \left(S_{q_1}^a\, S_{q_2}^a\right)\, \left(S_{q_3}^b\, S_{-q_1-q_2-q_3}^b\right) ,  \nonumber
\end{gather}
where $g({\bf q_1+q_2})= v_0+w_0|{\bf q_1 + q_2}|^{a-d}$ is the
Fourier transformation of the correlation function for the quenched random distribution of
defects in system. The Hamiltonian in (\ref{eq:HR}) is a functional of $n$ replications of the original
order parameter. The properties of the original disordered system after the RG
transformations of $H_R$ are obtained in the replica number limit
$n \to 0$.

The effective parameter of interaction $\tilde{u}_0=u_0-2g_0^2$ arising in the Hamiltonian
(\ref{eq:HR}) from influence of spin-elastic interaction determined by the parameter $g_0$ can
take both positive and negative values. As a result, the Hamiltonian in (\ref{eq:HR}) describes
a phase transitions both second order and first order. The tricritical behavior is realized
in the system for $\tilde{u}_0=0$. Furthermore, as it was shown in
Ref.~\onlinecite{Prudnikov01}, the dependence of the effective interaction in (\ref{eq:HR}) determined
by the vertex $g_0^2 -\tilde{\mu}_0(P)$ on outside pressure $P$ can lead to the phase
transition of second order for $P>P_t$ and lead to the phase transition of first order for
$P<P_t$, where $P_t$ is a tricritical value of outside pressure. But we restrict ourselves in
this paper by consideration of the critical behavior only.

As is known, in the field-theory approach \cite{Amit} the asymptotic critical behavior
of systems in the fluctuation region is determined by the Callan-Symanzik
RG equation for the vertex parts of the irreducible Green's functions.
To calculate the $\beta$ functions and the critical exponents as scaling $\gamma$ functions
of the renormalized interaction vertices $\tilde{u}$, $\tilde{g}^2=g^2-\tilde{\mu}$, $v$,
and $w$ appearing in the RG equation, we used the method based on the
Feynman diagram technique and the renormalization procedure.\cite{Zinn-Justin}

\begin{table*}
\caption{Coefficients for the $\beta$-functions in Eq.~\protect{\ref{eq:beta}}
(the continuation of Table~\protect{\ref{tab:b1}}).}
\label{tab:b2}
\begin{tabular}{dddddddd}\hline\hline
\multicolumn{1}{c}{$a$}&\multicolumn{1}{c}{$b_8$}&\multicolumn{1}{c}{$b_9$}&\multicolumn{1}{c}{$b_{10}$}
&\multicolumn{1}{c}{$b_{11}$}&\multicolumn{1}{c}{$b_{12}$}&\multicolumn{1}{c}{$b_{13}$} &\multicolumn{1}{c}{$b_{14}$}\\ \hline
3.00 & 4.000\,000 &  8.851\,848 &   9.703\,704 &   5.703\,704 &    -0.851\,848 &     1.703\,704  &     3.407\,408 \\
2.90 & 3.847\,968 &  8.018\,848 &   9.149\,424 &   5.384\,984 &    -0.630\,048 &     1.471\,544  &     2.943\,084 \\
2.80 & 3.777\,176 &  7.384\,568 &   8.671\,816 &   5.107\,714 &    -0.442\,952 &     1.259\,744  &     2.519\,492 \\
2.70 & 3.785\,160 &  6.909\,368 &   8.260\,288 &   4.866\,124 &    -0.278\,072 &     1.063\,416  &     2.126\,828 \\
2.60 & 3.875\,560 &  6.565\,360 &   7.906\,552 &   4.655\,620 &    -0.125\,208 &     0.878\,440  &     1.756\,888 \\
2.50 & 4.059\,204 &  6.332\,944 &   7.604\,032 &   4.472\,486 &    0.024\,984  &     0.701\,240  &     1.402\,484 \\
2.40 & 4.356\,584 &  6.198\,408 &   7.347\,528 &   4.313\,686 &    0.182\,264  &     0.528\,528  &     1.057\,056 \\
2.30 & 4.802\,876 &  6.152\,008 &   7.132\,944 &   4.176\,734 &    0.358\,480  &     0.357\,160  &     0.714\,324 \\
2.20 & 5.457\,744 &  6.185\,952 &   6.957\,112 &   4.059\,572 &    0.570\,112  &     0.184\,008  &     0.368\,020 \\
2.10 & 6.425\,424 &  6.291\,216 &   6.817\,672 &   3.960\,514 &    0.842\,960  &     0.005\,800  &     0.011\,596 \\
2.00 & 7.899\,532 &  6.451\,000 &   6.713\,000 &   3.878\,172 &    1.222\,400  &    -0.181\,032  &    -0.362\,064 \\ \hline\hline
\end{tabular}
\end{table*}

\begin{table*}
\caption{Coefficients for the $\gamma$-functions in
Eqs.~(\protect{\ref{eq:beta}}).}
\label{tab:c}
\begin{tabular}{ddddd}\hline\hline
\multicolumn{1}{c}{$a$}&\multicolumn{1}{c}{$c_1$}&\multicolumn{1}{c}{$c_2$}&\multicolumn{1}{c}{$c_3$}
&\multicolumn{1}{c}{$c_4$} \\  \hline
3.00  & 0.296\,288     & 0.592\,592    &   2.080\,000  & 4.000\,000  \\
2.90  & 0.423\,680     & 0.730\,760    &   1.683\,744  & 3.673\,840  \\
2.80  & 0.539\,712     & 0.874\,448    &   1.425\,600  & 3.393\,936  \\
2.70  & 0.651\,712     & 1.025\,656    &   1.207\,872  & 3.152\,488  \\
2.60  & 0.765\,696     & 1.186\,608    &   1.016\,128  & 2.943\,496  \\
2.50  & 0.887\,040     & 1.359\,760    &   0.837\,056  & 2.762\,240   \\
2.40  & 1.021\,088     & 1.547\,944    &   0.656\,544  & 2.605\,000   \\
2.30  & 1.173\,728     & 1.754\,448    &   0.456\,736  & 2.468\,768   \\
2.20  & 1.351\,808     & 1.983\,136    &   0.211\,584  & 2.351\,160   \\
2.10  & 1.563\,808     & 2.238\,656    &  -0.122\,112  & 2.250\,248   \\
2.00  & 1.820\,576     & 2.526\,584    &  -0.624\,224  & 2.164\,520   \\ \hline\hline
\end{tabular}
\end{table*}

As a result, we obtained the $\beta$ and $\gamma$ functions in the two-loop approximation
in the form of the expansion series in renormalized vertices $\tilde{u}$, $\tilde{g}^2$, $v$,
and $w$. We list here the resulting expansions:
\begin{eqnarray}
\displaystyle \beta_{\tilde{u}}(\tilde{u},v,w)= & - & \tilde{u}+9\tilde{u}^2-24\tilde{u}v-16(3f_1-f_2)\tilde{u}w   \nonumber \\
\displaystyle & - & \frac{308}{9}\tilde{u}^3-\frac{5920}{27}\tilde{u}v^2 +\frac{1664}{9}\tilde{u}^2v \nonumber \\
\displaystyle & + & 16(b_1+b_2)\tilde{u}^2w-64b_3\tilde{u}w^2-64b_4\tilde{u}vw, \nonumber \\
\displaystyle \beta_{\tilde{g}}(\tilde{u},\tilde{g},v,w)= & - & \tilde{g}^2+2\tilde{g}^4+6\tilde{u}\tilde{g}^2-8v\tilde{g}^2-16f_1w\tilde{g}^2   \nonumber \\
\displaystyle & - & \frac{92}{9}\tilde{u}^2\tilde{g}^2+\frac{224}{9}\tilde{u}v\tilde{g}^2 -\frac{736}{27}v^2\tilde{g}^2 \nonumber \\
\displaystyle & - & \frac{736}{27}b_5w^2\tilde{g}^2+\frac{112}{9}b_6\tilde{u}w\tilde{g}^2-\frac{736}{27}b_6vw\tilde{g}^2, \nonumber \\
\displaystyle \beta_v(\tilde{u},v,w)= & - & v-16v^2-16f_3w^2-16(3f_1-f_2)vw \nonumber \\
\displaystyle & + & 6\tilde{u}v -\frac{3040}{27}v^3-32b_7w^3 +\frac{800}{9}\tilde{u}v^2\nonumber \\
\displaystyle & + &12b_8\tilde{u}w^2 -32b_9vw^2-\frac{92}{9}\tilde{u}^2v \nonumber \\
\displaystyle & - & 32b_{10}v^2w+24b_{11}\tilde{u}vw, \nonumber \\
\displaystyle \beta_w(\tilde{u},v,w) & = & -(4-a)w-16(f_1-f_2)w^2-8vw \nonumber \\
\displaystyle & + & 6\tilde{u}w+32b_{12}w^3-32b_{13}vw^2 -\frac{736}{27}v^2w\nonumber \\
\displaystyle & - & \frac{92}{9}\tilde{u}^2w+12b_{14}\tilde{u}w^2 +\frac{368}{9}\tilde{u}vw, \nonumber
\end{eqnarray}
\begin{eqnarray}
\displaystyle \gamma_{\phi}(\tilde{u},v,w)= &  & 8f_2w+\frac{8}{9}\tilde{u}^2+\frac{64}{27}v^2+8c_{1}w^2 \nonumber\\
\displaystyle & - & \frac{32}{9}\tilde{u}v-6c_{2}\tilde{u}w+8c_{2}vw, \nonumber \\
\displaystyle \gamma_{\phi^2}(\tilde{u},\tilde{g},v,w)= & - & 3\tilde{u}+4v -2\tilde{g}^2 +8f_1w+6\tilde{u}^2 +16v^2 \nonumber \\
\displaystyle & + & 2\tilde{g}^4+8c_{3}w^2+8c_{4}vw -6c_{4}\tilde{u}w -24\tilde{u}v, \nonumber \\
 \displaystyle f_1 & = & \frac{(a-2)(a-4)}{2\sin(\pi a/2)}, \nonumber \\
   f_2&=&\frac{(a-2)(a-3)(a-4)}{48\pi \sin[\pi(a/2-1)]}, \nonumber \\
   f_3&=&\frac{(2a-5)(2a-7)}{2\sin[\pi(a-3/2)]}, \label{eq:beta}
\end{eqnarray}
where the coefficients $b_i$ and $c_i$ for different values of parameter $a$
in the range $2\leq a \leq 3$ are given in Tables~\ref{tab:b1}-\ref{tab:c}.
In the series (\ref{eq:beta}) it was used a standard change in variables
$\tilde{u} \to \tilde{u}/J$, $\tilde{g}^2 \to \tilde{g}^2/J$, $v \to v/J$, and $w \to w/J$,
where $J=\int {\rm d^{d}}q/(q^2+1)^2$ is the one-loop integral.

\begin{table*}
\caption{Stable fixed points and critical exponents for the compressible Ising
model with LR-correlated disorder.}
\label{tab:fp}
\begin{ruledtabular}
\begin{tabular}{ddddddddd}
\multicolumn{1}{c}{$a$} & \multicolumn{1}{c}{$\tilde{u}^*$} & \multicolumn{1}{c}{$\tilde{g}^{*2}$}  & \multicolumn{1}{c}{$v^*$}  & \multicolumn{1}{c}{$v^*+w^*$}
& \multicolumn{1}{c}{$\nu$}    & \multicolumn{1}{c}{$\alpha$}     & \multicolumn{1}{c}{$\beta$} & \multicolumn{1}{c}{$z$}       \\  \hline
 3.01  & 0.26482 & 0.02043 & 0.03448 & 0.03448 & 0.6896 & -0.0687 & 0.3561 & 2.1712  \\
 3.00  & 0.26482 & 0.02043 & 0.01393 & 0.03448 & 0.6896 & -0.0687 & 0.3561 & 2.1712  \\
 2.90  & 0.28867 & 0.06021 & 0.01993 & 0.04257 & 0.7387 & -0.2160 & 0.3806 & 2.2120  \\
 2.80  & 0.30881 & 0.05443 & 0.02510 & 0.04894 & 0.7411 & -0.2232 & 0.3805 & 2.2486  \\
 2.70  & 0.32670 & 0.04708 & 0.02968 & 0.05422 & 0.7402 & -0.2205 & 0.3785 & 2.2837  \\
 2.60  & 0.34294 & 0.03847 & 0.03380 & 0.05870 & 0.7369 & -0.2106 & 0.3749 & 2.3184  \\
 2.50  & 0.35776 & 0.02899 & 0.03752 & 0.06248 & 0.7320 & -0.1958 & 0.3703 & 2.3532  \\
 2.40  & 0.37120 & 0.02281 & 0.04086 & 0.06562 & 0.7297 & -0.1890 & 0.3668 & 2.3879  \\
 2.30  & 0.38313 & 0.01772 & 0.04380 & 0.06811 & 0.7278 & -0.1833 & 0.3634 & 2.4215  \\
 2.20  & 0.39322 & 0.01281 & 0.04631 & 0.06989 & 0.7254 & -0.1761 & 0.3597 & 2.4524  \\
 2.10  & 0.40090 & 0.00762 & 0.04829 & 0.07084 & 0.7219 & -0.1657 & 0.3557 & 2.4780  \\
 2.00  & 0.40521 & 0.00186 & 0.04959 & 0.07074 & 0.7169 & -0.1507 & 0.3511 & 2.4949  \\
\end{tabular}
\end{ruledtabular}
\end{table*}

We should like note that WH model with $a=3$ corresponds
to a system with point-like defects. The case with $a=2$
corresponds to a system of straight lines of impurities
or straight dislocation lines of random orientations
in a sample. The cases with noninteger $a$ with $2<a<3$
are treated in terms of a fractal dimension of impurities.
We think that the features of the critical behavior of
disordered systems with $2<a<3$ can be displayed in real
highly disordered systems when concentration of defects
is sufficiently high that effective long-range correlations
can occur because of elastic interaction of defects.
For values of correlation parameter a less than $1.5$,
the unavoidable divergencies appear in diagrams for
RG $\beta$- and $\gamma$-functions in framework
of theoretical-field approach with fixed dimension $d=3$,
which is used in this paper.

\section{FP's and various types of critical behavior. Critical exponents}

The nature of the critical behavior is determined by the existence of a stable FP satisfying
the system of equations
\begin{equation}
\label{fp}
\beta_i(\tilde{u}^*,\tilde{g}^*,v^*,w^*)=0 \ \ \ (i=1,2,3,4).
\end{equation}
It is well known that perturbation series are asymptotic series, and that the vertices
describing the interaction of the order parameter fluctuations in the fluctuating region
$\tau \to 0$ are large enough so that expressions (\ref{eq:beta}) cannot be used
directly. For this reason, to extract the required physical information
from the obtained expressions, we employed the Pad\'{e}-Borel approximation of the summation
of asymptotic series extended to the multiparameter case.\cite{PrudnikovPR00}
The direct and the inverse Borel transformations for the multiparameter case
have the form
\begin{eqnarray}
\label{rl}
  \displaystyle  f(u_1,u_2,u_3,u_4)&=&\sum\limits_{i,j,l,k}c_{ijlk}u_1^i u_2^j u_3^lu_4^k= \nonumber \\
  &=&\int\limits_{0}^{\infty}e^{-t}B(u_1t,u_2t,u_3t,u_4t)dt,  \\
  \displaystyle B(u_1,u_2,u_3,u_4)&=&\sum\limits_{i,j,l,k}\displaystyle\frac{c_{ijlk}}{(i+j+l+k)!}\,u_1^i u_2^j u_3^lu_4^k. \nonumber
\end{eqnarray}
A series in the auxiliary variable $\theta$ is introduced for analytical
continuation of the Borel transform of the function:
\begin{eqnarray}
   & &{\tilde{B}}(u_1,u_2,u_3,u_4,\theta)= \\
   & &=\sum\limits_{k=0}^{\infty}\theta^k\sum\limits_{i=0}^{k}\sum\limits_{j=0}^{k-i}\sum\limits_{l=0}^{k-i-j}\frac{c_{i,j,\,l,\, k-i-j-l}}{k!}u_1^i u_2^j u_3^lu_4^{k-i-j-l}, \nonumber
\end{eqnarray}
to which the [L/M] Pad\'{e} approximation is applied at the point $\theta=1$.
To perform the analytical continuation, the Pad\'{e} approximant of [L/1] type may be
used which is known to provide rather good results for various Landau-Wilson models
(see, e.g., Refs.~\onlinecite{Sokolov} and \onlinecite{Baker}). The property of preserving
the symmetry of a system during application of the Pad\'{e}
approximation by the $\theta$ method, as in Ref.~\onlinecite{Sokolov}, has become
important for multivertices models. We used the [2/1] approximant to calculate
the $\beta$ functions in the two-loop approximation.

However, analysis of the series coefficients for the $\beta_w$ function has shown that
the summation of this series is fairly poor, which resulted in the absence of FP's with
$w^* \neq 0$ for $a<2.93$. Dorogovtsev \cite{Dorogovtsev} found the symmetry of the scaling
function for the WH model in relation to the transformation $(u,v,w) \to (u,v,v+w)$,
which gives the possibility of investigating the problem of the existence of FP's
with $w^* \neq 0$ in the variables $(u,v,v+w)$. In this case, our investigations
carried out in Ref.\onlinecite{PrudnikovPR00} have shown the existence of FP's with
$w^* \neq 0$ in the whole region where the parameter $a$ changes.

We have found two classes of FP's with $g^*=0$ for an incompressible Ising model and
with $g^*>0$ for the compressible Ising model. If a possibility of realization of
multicritical behavior in system for $\tilde{u}^*=u^*-2g^{*2}=0$ or for
$\tilde{g}^{*2}=g^{*2}-\tilde{\mu}^*=0$ does not consider in this paper
(for critical behavior \cite{Prudnikov01} $\tilde{\mu}^*=0$ and $\tilde{g}^{*}=g^{*}$), then for
either of the two classes there are three types of FP's in the physical region of parameter
space, $\tilde{u}^*, g^{*2}, v^*$, and $v^*+w^*>0$ for different values of $a$.
Type I corresponds to the FP of a pure system ($\tilde{u}^* \neq 0$,$v^*,w^*=0$),
type II is a SR-disorder FP ($\tilde{u}^*,v^* \neq 0$, $w^*=0$), and type III corresponds
to LR-disorder FP ($\tilde{u}^*,v^*,w^* \neq 0$).

The type of critical behavior of this disordered system for each value of $a$ is determined
by the stability of the corresponding FP. The stability properties of the FP's are controlled
by the eigenvalues $\lambda_i$ of the matrix
\begin{equation}
\label{stab}
\Omega_{ij}=\frac{\partial \beta_i}{\partial u_j}
\end{equation}
computed at the given FP: a FP is stable if all eigenvalues $\lambda_i$ are positive.
If some of the eigenvalues $\lambda_i$ are complex numbers, then the real parts of these
eigenvalues must be positive for stable FP.

Our calculations showed that FP's which belong to the class of incompressible Ising model
with $g^*=0$ are unstable with respect to the influence of elastic deformations.
For class of FP's corresponding to the compressible Ising model with $g^* \neq 0$
the values of the stable FP's obtained for $2\leq a \leq 3$ are presented in Table~\ref{tab:fp}.
As one can see from this Table, for the compressible Ising model the LR-disorder FP is
stable for values of $a$ in the whole investigated range. The additional calculations
for $3<a<4$ have shown that only SR-disorder FP is stable in this range. For $a=3$ FP values
for vertices $\tilde{u}$ and $g(k)$ are equal, $\tilde{u}^*=0.26482$ and $v^*+w^*=0.03448$,
and correspond to the SR-disorder FP for compressible Ising model, although $w^* \neq 0$.
Similarly, for $a=3$ the LR disorder is marginal, and the critical behavior of the
compressible Ising model with LR-correlated disorder, as that of the SR-disordered
compressible Ising model, is characterized by the same critical exponents (Table~\ref{tab:fp}).

We have calculated the static critical exponents $\eta$ and $\nu$ for the compressible Ising
model with LR-correlated disorder (Table~\ref{tab:fp}), received from the resummed by the
generalized Pad\'{e}-Borel method $\gamma$ functions in the corresponding stable FP's:
\begin{eqnarray}
\label{exps}
\eta&=&\gamma_{\phi}(\tilde{u}^*,\tilde{g}^*,v^*,w^*), \\
\nu&=&[2+\gamma_{\phi^2}(\tilde{u}^*,\tilde{g}^*,v^*,w^*)-\gamma_{\phi}(\tilde{u}^*,\tilde{g}^*,v^*,w^*)]^{-1}. \nonumber
\end{eqnarray}
Values of the specific-heat exponent $\alpha$ and the exponent $\beta$ for magnetization
which are required for calculation of ultrasound characteristics were determined from
scaling relations $\alpha=2-d\nu$ and $\beta=(d-2+\eta)\nu/2$ for
$2\leq a \leq 3$ and presented in Table~\ref{tab:fp}.
The values of the dynamic critical exponent $z$ also presented in Table~\ref{tab:fp}
for $2\leq a \leq 3$ were taken from our paper,\cite{PrudnikovPR00} where the critical
dynamics of 3D disordered Ising model with LR-correlated disorder was considered for purely
relaxational model A.
It is caused by fact \cite{HH} that the coupling of the order parameter with elastic
deformations is irrelevant for the relaxational critical properties of the order parameter
in disordered compressible Ising model with a negative specific-heat exponent $\alpha$.

\begin{figure}
\includegraphics[width=0.45\textwidth]{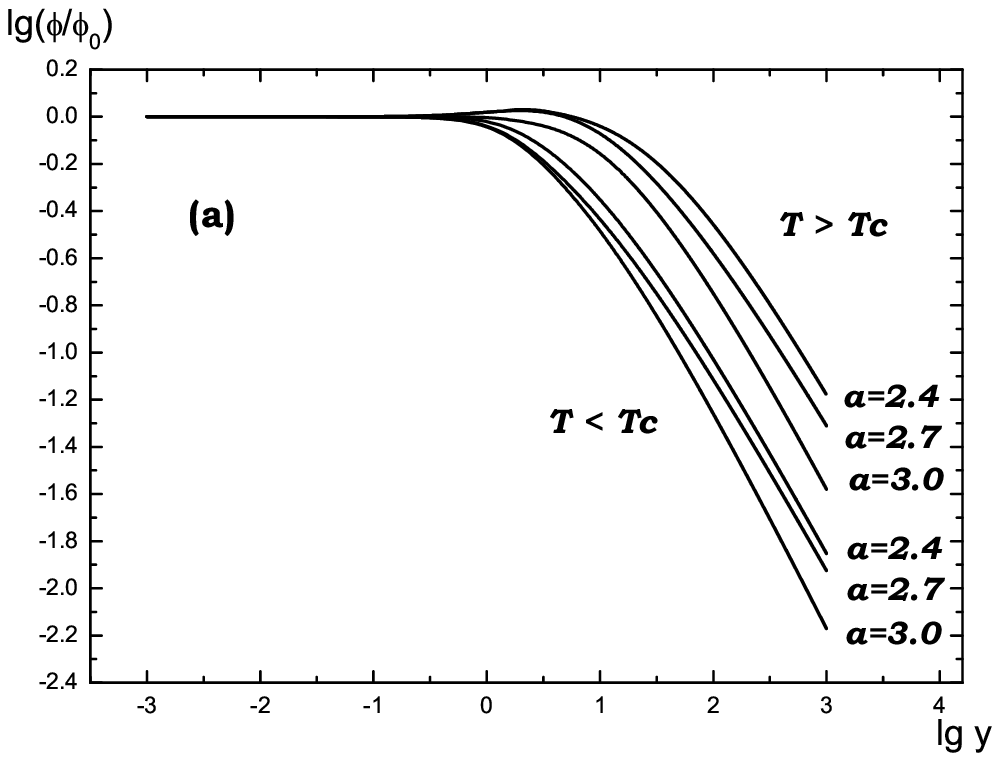}
\includegraphics[width=0.45\textwidth]{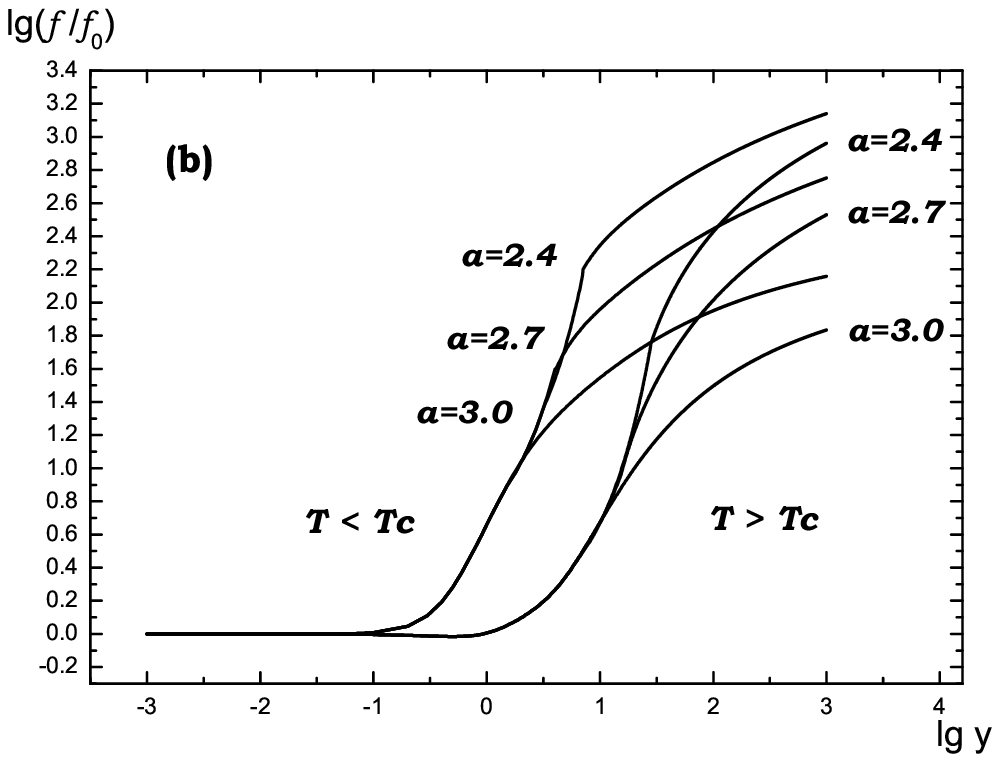}
\caption{ \label{fig:2} \textbf{(a)} Scaling functions for the critical sound attenuation $\phi(y)$
and \textbf{(b)} sound velocity dispersion $f(y)$ in a double-logarithmic plot for disordered systems characterized by different values
of the correlation parameter $a$ at $T>T_c$ and $T<T_c$ [$\phi_0=\phi(0)$, $f_0=f(0)$]. }
\end{figure}
\section{Analysis of results for ultrasound critical characteristics and conclusions}

In this work, we employed the Pad\'{e}-Borel approximation of the summation
of asymptotic series for calculation of the ultrasound scaling functions $\phi (y)$
 in (\ref{ScaleFf}) using approximant [1/1].
The short series for scaling functions $\phi (y)$ were summed on the values of
vertices $u$, $v$, and $w$ in LR-disorder FP for $2\leq a \leq 3$
with variable $y$ which was changed in the interval from $10^{-3}$ to $10^{3}$ by
steps with $\Delta y=0.1$.

The resultant behavior of the dynamic scaling functions $\phi(y)$ and $f(y)$
for individual values of $a$ is shown in Figs.~\ref{fig:2}(a) and~\ref{fig:2}(b)
on a log-log scale.
Depending on the interval of changing variable $y$, the following
asymptotic regions can be distinguished in the behavior of $\phi(y)$:
a hydrodynamic region, where $y \sim \omega \xi^z \sim
\left( q \xi \right)^z \ll 1 $, and a critical region $y \sim \omega
\xi^z \gg 1$, which determines the behavior of the system near the
phase transition temperature ($\tau=(T-T_c)/T_c \ll 1$). We have seen
from these curves that the correlation properties of a structural disorder
does not affect the behavior of the scaling functions $\phi(y)$
in the hydrodynamic region with $y\ll 1$.
However, LR-disorder begins to manifest itself in the crossover region
$10^{-1} < y < 10$ and it has a drastic effect in the critical region with $y>10$
($T \to T_c$).

As follows from Eqs.~(\ref{Atten}) and (\ref{ImScF}) , the
attenuation coefficient can be expressed as
\begin{equation}
\label{ass:1}
\alpha\left(\omega,\tau\right)\sim\omega^{2}\tau^{-\alpha-\nu{z}}\phi\left(y\right),
\end{equation}
and, using Eqs.~(\ref{disp}) and (\ref{ReScF}), we can write the
relation for the sound velocity dispersion in the form
\begin{equation}
\label{ass:2}
    c^{2}(\omega,\tau) - c^{2}(0,\tau) \sim \tau^{-\alpha}\left[f(y)-f(0)\right].
\end{equation}

The results of the calculation of the asymptotic dependences of the
attenuation coefficient and the sound velocity dispersion for the
critical and hydrodynamic regions are given in the Table~\ref{tab:ultr}. The
characteristics of their frequency and temperature dependences were
determined in the range $10^{-3}\leq y \leq 10^{-1}$ for the
hydrodynamic regime and in the range $10\leq y \leq 10^3$ for the
critical regime. Note that, according to,\cite{Matching} the real
temperature range $10^{-3}\leq \tau \leq 10^{-1}$ in ultrasonic
studies of phase transitions corresponds to the range $1\leq y \leq
10^2$, i.e., it covers the crossover region and the beginning of the
critical region (precritical regime).

\begin{table*}
\caption{Asymptotic behavior of the sound attenuation coefficient $\alpha(\omega,\tau)$
in the critical, precritical, and hydrodynamic regimes for pure and disordered systems}
\label{tab:ultr}
\begin{ruledtabular}
\begin{tabular}{llll|lll}
        & \multicolumn{3}{c|}{$T<T_{c}$}                                                           & \multicolumn{3}{c}{$T>T_{c}$}  \\ \hline \rule{0pt}{4mm}
 System & \multicolumn{1}{c}{Hydrodynamic}    & \multicolumn{1}{c}{Precritical}     & \multicolumn{1}{c|}{Critical}        & \multicolumn{1}{c}{Hydrodynamic}    & \multicolumn{1}{c}{Precritical}     & \multicolumn{1}{c}{Critical}       \\ \hline \rule{0pt}{5mm}
 Pure   & $ \omega^{\displaystyle 2.00}\,\tau^{\displaystyle -1.38 }$ & $ \omega^{\displaystyle 1.08}\,\tau^{\displaystyle -0.21 }$ & $ \omega^{\displaystyle 0.98}\,\tau^{\displaystyle -0.08 }$ & $ \omega^{\displaystyle 2.00}\,\tau^{\displaystyle -1.38 }$ & $ \omega^{\displaystyle 1.20}\,\tau^{\displaystyle -0.37 }$ & $ \omega^{\displaystyle 1.05}\,\tau^{\displaystyle -0.17 }$ \\
$a=3.0$ & ${\omega^{\displaystyle 2.00}\,\tau^{\displaystyle -1.43}}$ & ${\omega^{\displaystyle 1.21}\,\tau^{\displaystyle -0.24}}$ & ${\omega^{\displaystyle 1.11}\,\tau^{\displaystyle -0.10}}$ & ${\omega^{\displaystyle 2.00}\,\tau^{\displaystyle -1.43}}$ & ${\omega^{\displaystyle 1.37}\,\tau^{\displaystyle -0.49}}$ & ${\omega^{\displaystyle 1.21}\,\tau^{\displaystyle -0.24}}$ \\
$a=2.9$ & ${\omega^{\displaystyle 2.00}\,\tau^{\displaystyle -1.43}}$ & ${\omega^{\displaystyle 1.30}\,\tau^{\displaystyle -0.27}}$ & ${\omega^{\displaystyle 1.21}\,\tau^{\displaystyle -0.12}}$ & ${\omega^{\displaystyle 2.00}\,\tau^{\displaystyle -1.43}}$ & ${\omega^{\displaystyle 1.46}\,\tau^{\displaystyle -0.54}}$ & ${\omega^{\displaystyle 1.29}\,\tau^{\displaystyle -0.28}}$ \\
$a=2.8$ & ${\omega^{\displaystyle 2.00}\,\tau^{\displaystyle -1.44}}$ & ${\omega^{\displaystyle 1.30}\,\tau^{\displaystyle -0.28}}$ & ${\omega^{\displaystyle 1.21}\,\tau^{\displaystyle -0.13}}$ & ${\omega^{\displaystyle 2.00}\,\tau^{\displaystyle -1.44}}$ & ${\omega^{\displaystyle 1.46}\,\tau^{\displaystyle -0.55}}$ & ${\omega^{\displaystyle 1.29}\,\tau^{\displaystyle -0.29}}$ \\
$a=2.7$ & ${\omega^{\displaystyle 2.00}\,\tau^{\displaystyle -1.47}}$ & ${\omega^{\displaystyle 1.30}\,\tau^{\displaystyle -0.29}}$ & ${\omega^{\displaystyle 1.21}\,\tau^{\displaystyle -0.13}}$ & ${\omega^{\displaystyle 2.00}\,\tau^{\displaystyle -1.47}}$ & ${\omega^{\displaystyle 1.46}\,\tau^{\displaystyle -0.56}}$ & ${\omega^{\displaystyle 1.29}\,\tau^{\displaystyle -0.29}}$ \\
$a=2.6$ & ${\omega^{\displaystyle 2.00}\,\tau^{\displaystyle -1.50}}$ & ${\omega^{\displaystyle 1.30}\,\tau^{\displaystyle -0.30}}$ & ${\omega^{\displaystyle 1.21}\,\tau^{\displaystyle -0.14}}$ & ${\omega^{\displaystyle 2.00}\,\tau^{\displaystyle -1.50}}$ & ${\omega^{\displaystyle 1.46}\,\tau^{\displaystyle -0.57}}$ & ${\omega^{\displaystyle 1.30}\,\tau^{\displaystyle -0.30}}$ \\
$a=2.5$ & ${\omega^{\displaystyle 2.00}\,\tau^{\displaystyle -1.53}}$ & ${\omega^{\displaystyle 1.30}\,\tau^{\displaystyle -0.32}}$ & ${\omega^{\displaystyle 1.21}\,\tau^{\displaystyle -0.15}}$ & ${\omega^{\displaystyle 2.00}\,\tau^{\displaystyle -1.53}}$ & ${\omega^{\displaystyle 1.46}\,\tau^{\displaystyle -0.59}}$ & ${\omega^{\displaystyle 1.30}\,\tau^{\displaystyle -0.30}}$ \\
$a=2.4$ & ${\omega^{\displaystyle 2.00}\,\tau^{\displaystyle -1.55}}$ & ${\omega^{\displaystyle 1.30}\,\tau^{\displaystyle -0.34}}$ & ${\omega^{\displaystyle 1.21}\,\tau^{\displaystyle -0.16}}$ & ${\omega^{\displaystyle 2.00}\,\tau^{\displaystyle -1.55}}$ & ${\omega^{\displaystyle 1.47}\,\tau^{\displaystyle -0.63}}$ & ${\omega^{\displaystyle 1.30}\,\tau^{\displaystyle -0.32}}$ \\
$a=2.3$ & ${\omega^{\displaystyle 2.00}\,\tau^{\displaystyle -1.58}}$ & ${\omega^{\displaystyle 1.32}\,\tau^{\displaystyle -0.37}}$ & ${\omega^{\displaystyle 1.21}\,\tau^{\displaystyle -0.17}}$ & ${\omega^{\displaystyle 2.00}\,\tau^{\displaystyle -1.58}}$ & ${\omega^{\displaystyle 1.51}\,\tau^{\displaystyle -0.71}}$ & ${\omega^{\displaystyle 1.30}\,\tau^{\displaystyle -0.34}}$ \\
$a=2.2$ & ${\omega^{\displaystyle 2.00}\,\tau^{\displaystyle -1.60}}$ & ${\omega^{\displaystyle 1.32}\,\tau^{\displaystyle -0.39}}$ & ${\omega^{\displaystyle 1.22}\,\tau^{\displaystyle -0.19}}$ & ${\omega^{\displaystyle 2.00}\,\tau^{\displaystyle -1.60}}$ & ${\omega^{\displaystyle 1.53}\,\tau^{\displaystyle -0.76}}$ & ${\omega^{\displaystyle 1.31}\,\tau^{\displaystyle -0.37}}$ \\
$a=2.1$ & ${\omega^{\displaystyle 2.00}\,\tau^{\displaystyle -1.62}}$ & ${\omega^{\displaystyle 1.33}\,\tau^{\displaystyle -0.42}}$ & ${\omega^{\displaystyle 1.23}\,\tau^{\displaystyle -0.21}}$ & ${\omega^{\displaystyle 2.00}\,\tau^{\displaystyle -1.62}}$ & ${\omega^{\displaystyle 1.56}\,\tau^{\displaystyle -0.83}}$ & ${\omega^{\displaystyle 1.32}\,\tau^{\displaystyle -0.41}}$ \\
$a=2.0$ & ${\omega^{\displaystyle 2.00}\,\tau^{\displaystyle -1.64}}$ & ${\omega^{\displaystyle 1.34}\,\tau^{\displaystyle -0.45}}$ & ${\omega^{\displaystyle 1.24}\,\tau^{\displaystyle -0.23}}$ & ${\omega^{\displaystyle 2.00}\,\tau^{\displaystyle -1.64}}$ & ${\omega^{\displaystyle 1.60}\,\tau^{\displaystyle -0.92}}$ & ${\omega^{\displaystyle 1.33}\,\tau^{\displaystyle -0.44}}$ \\
\end{tabular}
\end{ruledtabular}
\end{table*}

\begin{table*}
\caption{Asymptotic behavior of the sound velocity dispersion $c^2(\omega,\tau)-c^2(0,\tau)$
in the critical, precritical, and hydrodynamic regimes for pure and disordered systems }
\label{tab:disp}
\begin{ruledtabular}
\begin{tabular}{llll|lll}
        & \multicolumn{3}{c|}{$T<T_{c}$}                                                           & \multicolumn{3}{c}{$T>T_{c}$}  \\ \hline \rule{0pt}{4mm}
 System & \multicolumn{1}{c}{Hydrodynamic}    & \multicolumn{1}{c}{Precritical}     & \multicolumn{1}{c|}{Critical}        & \multicolumn{1}{c}{Hydrodynamic}    & \multicolumn{1}{c}{Precritical}     & \multicolumn{1}{c}{Critical}       \\ \hline \rule{0pt}{5mm}
 Pure   & $ \omega^{\displaystyle 2.00}\,\tau^{\displaystyle -2.65 }$ & $ \omega^{\displaystyle 0.30}\,\tau^{\displaystyle -0.49 }$ & $ \omega^{\displaystyle 0.11}\,\tau^{\displaystyle -0.25 }$ & $ \omega^{\displaystyle 2.00}\,\tau^{\displaystyle -2.65 }$ & $ \omega^{\displaystyle 1.08}\,\tau^{\displaystyle -1.48 }$ & $ \omega^{\displaystyle 0.34}\,\tau^{\displaystyle -0.54 }$ \\
$a=3.0$ & ${\omega^{\displaystyle 2.00}\,\tau^{\displaystyle -2.93}}$ & ${\omega^{\displaystyle 0.38 }\,\tau^{\displaystyle -0.51 }}$ & ${\omega^{\displaystyle 0.25 }\,\tau^{\displaystyle -0.30 }}$ & ${\omega^{\displaystyle 2.00}\,\tau^{\displaystyle -2.93}}$ & ${\omega^{\displaystyle 1.17 }\,\tau^{\displaystyle -1.68 }}$ & ${\omega^{\displaystyle 0.46 }\,\tau^{\displaystyle -0.62 }}$ \\
$a=2.9$ & ${\omega^{\displaystyle 2.00}\,\tau^{\displaystyle -3.05}}$ & ${\omega^{\displaystyle 0.46 }\,\tau^{\displaystyle -0.54 }}$ & ${\omega^{\displaystyle 0.32 }\,\tau^{\displaystyle -0.33 }}$ & ${\omega^{\displaystyle 2.00}\,\tau^{\displaystyle -3.05}}$ & ${\omega^{\displaystyle 1.22 }\,\tau^{\displaystyle -1.77 }}$ & ${\omega^{\displaystyle 0.57 }\,\tau^{\displaystyle -0.72 }}$ \\
$a=2.8$ & ${\omega^{\displaystyle 2.00}\,\tau^{\displaystyle -3.11}}$ & ${\omega^{\displaystyle 0.46 }\,\tau^{\displaystyle -0.55 }}$ & ${\omega^{\displaystyle 0.33 }\,\tau^{\displaystyle -0.34 }}$ & ${\omega^{\displaystyle 2.00}\,\tau^{\displaystyle -3.11}}$ & ${\omega^{\displaystyle 1.32 }\,\tau^{\displaystyle -1.98 }}$ & ${\omega^{\displaystyle 0.61 }\,\tau^{\displaystyle -0.80 }}$ \\
$a=2.7$ & ${\omega^{\displaystyle 2.00}\,\tau^{\displaystyle -3.16}}$ & ${\omega^{\displaystyle 0.46 }\,\tau^{\displaystyle -0.56 }}$ & ${\omega^{\displaystyle 0.33 }\,\tau^{\displaystyle -0.35 }}$ & ${\omega^{\displaystyle 2.00}\,\tau^{\displaystyle -3.16}}$ & ${\omega^{\displaystyle 1.42 }\,\tau^{\displaystyle -2.18 }}$ & ${\omega^{\displaystyle 0.65 }\,\tau^{\displaystyle -0.88 }}$ \\
$a=2.6$ & ${\omega^{\displaystyle 2.00}\,\tau^{\displaystyle -3.21}}$ & ${\omega^{\displaystyle 0.46 }\,\tau^{\displaystyle -0.58 }}$ & ${\omega^{\displaystyle 0.33 }\,\tau^{\displaystyle -0.36 }}$ & ${\omega^{\displaystyle 2.00}\,\tau^{\displaystyle -3.21}}$ & ${\omega^{\displaystyle 1.49 }\,\tau^{\displaystyle -2.34 }}$ & ${\omega^{\displaystyle 0.66 }\,\tau^{\displaystyle -0.92 }}$ \\
$a=2.5$ & ${\omega^{\displaystyle 2.00}\,\tau^{\displaystyle -3.25}}$ & ${\omega^{\displaystyle 0.46 }\,\tau^{\displaystyle -0.60 }}$ & ${\omega^{\displaystyle 0.33 }\,\tau^{\displaystyle -0.37 }}$ & ${\omega^{\displaystyle 2.00}\,\tau^{\displaystyle -3.25}}$ & ${\omega^{\displaystyle 1.58 }\,\tau^{\displaystyle -2.53 }}$ & ${\omega^{\displaystyle 0.66 }\,\tau^{\displaystyle -0.94 }}$ \\
$a=2.4$ & ${\omega^{\displaystyle 2.00}\,\tau^{\displaystyle -3.30}}$ & ${\omega^{\displaystyle 0.46 }\,\tau^{\displaystyle -0.62 }}$ & ${\omega^{\displaystyle 0.33 }\,\tau^{\displaystyle -0.38 }}$ & ${\omega^{\displaystyle 2.00}\,\tau^{\displaystyle -3.30}}$ & ${\omega^{\displaystyle 1.62 }\,\tau^{\displaystyle -2.64 }}$ & ${\omega^{\displaystyle 0.67 }\,\tau^{\displaystyle -0.97 }}$ \\
$a=2.3$ & ${\omega^{\displaystyle 2.00}\,\tau^{\displaystyle -3.34}}$ & ${\omega^{\displaystyle 0.47 }\,\tau^{\displaystyle -0.65 }}$ & ${\omega^{\displaystyle 0.33 }\,\tau^{\displaystyle -0.39 }}$ & ${\omega^{\displaystyle 2.00}\,\tau^{\displaystyle -3.34}}$ & ${\omega^{\displaystyle 1.66 }\,\tau^{\displaystyle -2.75 }}$ & ${\omega^{\displaystyle 0.67 }\,\tau^{\displaystyle -1.00 }}$ \\
$a=2.2$ & ${\omega^{\displaystyle 2.00}\,\tau^{\displaystyle -3.38}}$ & ${\omega^{\displaystyle 0.50 }\,\tau^{\displaystyle -0.70 }}$ & ${\omega^{\displaystyle 0.34 }\,\tau^{\displaystyle -0.39 }}$ & ${\omega^{\displaystyle 2.00}\,\tau^{\displaystyle -3.38}}$ & ${\omega^{\displaystyle 1.72 }\,\tau^{\displaystyle -2.89 }}$ & ${\omega^{\displaystyle 0.68 }\,\tau^{\displaystyle -1.03 }}$ \\
$a=2.1$ & ${\omega^{\displaystyle 2.00}\,\tau^{\displaystyle -3.41}}$ & ${\omega^{\displaystyle 0.51 }\,\tau^{\displaystyle -0.74 }}$ & ${\omega^{\displaystyle 0.35 }\,\tau^{\displaystyle -0.41 }}$ & ${\omega^{\displaystyle 2.00}\,\tau^{\displaystyle -3.41}}$ & ${\omega^{\displaystyle 1.76 }\,\tau^{\displaystyle -2.98 }}$ & ${\omega^{\displaystyle 0.70 }\,\tau^{\displaystyle -1.09 }}$ \\
$a=2.0$ & ${\omega^{\displaystyle 2.00}\,\tau^{\displaystyle -3.43}}$ & ${\omega^{\displaystyle 0.57 }\,\tau^{\displaystyle -0.87 }}$ & ${\omega^{\displaystyle 0.37 }\,\tau^{\displaystyle -0.51 }}$ & ${\omega^{\displaystyle 2.00}\,\tau^{\displaystyle -3.43}}$ & ${\omega^{\displaystyle 1.80 }\,\tau^{\displaystyle -3.06 }}$ & ${\omega^{\displaystyle 0.75 }\,\tau^{\displaystyle -1.18 }}$ \\
\end{tabular}
\end{ruledtabular}
\end{table*}

\begin{figure}[b]
\includegraphics[width=0.5\textwidth]{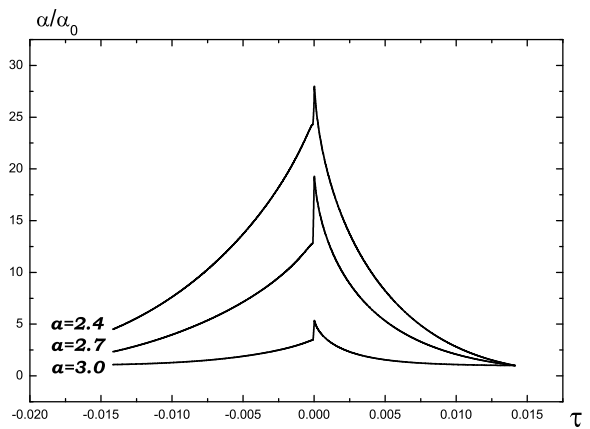}
\caption{ \label{fig:3} Temperature dependence of the attenuation
coefficient $\alpha(\omega,\tau)$ calculated for systems with LR-correlated disorder
characterized by different values of the correlation parameter $a$
at $B=0.3$ and $\omega/\Gamma_0=0.0015$.
}
\end{figure}

As it follows from the Table~\ref{tab:ultr}, the increase in correlation effects for
disorder characterized by decrease in parameter $a$ values leads to systematical
increase in the values of exponents $k_{\omega}^{(\alpha)(c)}$ and $k_{\tau}^{(\alpha)(c)}$,
which we can introduce for description of the critical anomalies of the frequency and temperature
dependences of the attenuation coefficient $\alpha \sim \omega^{k_{\omega}^{(\alpha)}}\tau^{-k_{\tau}^{(\alpha)}}$
and the sound velocity dispersion $c^{2}(\omega,\tau) - c^{2}(0,\tau) \sim \omega^{k_{\omega}^{(c)}}\tau^{-k_{\tau}^{(c)}}$.
So, for the attenuation coefficient the values of exponents $k_{\omega}^{(\alpha)}=1.21$
and $k_{\tau}^{(\alpha)}=0.24$ for $T>T_c$ and $k_{\omega}^{(\alpha)}=1.11$ and $k_{\tau}^{(\alpha)}=0.10$
for $T<T_c$ in the case of point-like uncorrelated defects characterized by parameter
$a \geq 3.0$, $k_{\omega}^{(\alpha)}=1.29$ and $k_{\tau}^{(\alpha)}=0.29$ for $T>T_c$ and
$k_{\omega}^{(\alpha)}=1.21$ and $k_{\tau}^{(\alpha)}=0.13$ for $T<T_c$ in the case of complex
structure defects characterized, for example, by value of the parameter $a= 2.7$, and
$k_{\omega}^{(\alpha)}=1.30$ and $k_{\tau}^{(\alpha)}=0.32$ for $T>T_c$ and
$k_{\omega}^{(\alpha)}=1.21$ and $k_{\tau}^{(\alpha)}=0.16$ for $T<T_c$ in the case of extended
linear defects characterized by value of the parameter $a= 2.4$. The comparison of these
exponent values with the values of $k_{\omega}^{(\alpha)}= 1.05$ and $k_{\tau}^{(\alpha)}=0.17$
for $T>T_c$ and $k_{\omega}^{(\alpha)}=0.98$ and $k_{\tau}^{(\alpha)}=0.08$ for $T<T_c$
calculated in our paper \cite{Prudnikov_JETP} for a pure Ising-like systems shows
the strong influence of disorder and its correlation effects on the frequency and temperature
dependences of the attenuation coefficient in the vicinity of the critical point.

We must note that the increase in the attenuation coefficient
with increase in LR correlations for disorder as the critical temperature
is approached is expected to be stronger than for systems with uncorrelated disorder
or for the pure systems even in the hydrodynamic region. At the same time in the critical region,
the systems with LR-correlated disorder should exhibit stronger both frequency and temperature
dependences of the attenuation coefficient than those for systems with uncorrelated
disorder or for pure systems.

The similar conclusions we can make in relation to the frequency and temperature
dependences of the sound velocity dispersion in the critical range on basis of those
values of exponents $k_{\omega}^{(c)}$ and $k_{\tau}^{(c)}$ which are given in Table~\ref{tab:disp}.
and the values of $k_{\omega}^{(c)}= 0.34$ and $k_{\tau}^{(c)}=0.54$
for $T>T_c$ and $k_{\omega}^{(c)}=0.11$ and $k_{\tau}^{(c)}=0.25$ for $T<T_c$
calculated in paper \cite{Prudnikov_JETP} for pure systems.

These conclusions are supported by the model representation of the
results of the numerical calculations of the attenuation coefficient
(Fig.~\ref{fig:3}) for systems with LR-correlated disorder
for different values of the parameter $a$ performed at $B=0.3$
and $\omega/\Gamma_0=0.0015$. These values were determined in
Ref.~\onlinecite{Prudnikov_JETP} when we compared the calculated
temperature dependence of the attenuation coefficient and the results
of experimental studies of pure $\mathrm{\mathop{FeF_2}}$ samples,\cite{IkushimaF}
which demonstrate Ising-like
behavior in the critical region.

A particularly important result of our investigation consists in the
predicted manifestation of the dynamical effects of structural defects
on anomalous ultrasound attenuation and dispersion over a
wider temperature range near the critical temperature (already in
the hydrodynamic region) in comparison with other experimental
methods,\cite{Rosov} which require a narrow temperature range (of
about $\tau \simeq 10^{-4}$) to be studied for revealing these
effects. Thus, the results obtained in this paper can serve as a reference point
for purposeful experimental investigations of the dynamical effects of
structural defects and their correlation properties on the critical
behavior of solids using acoustic methods via the detection of the peculiarities
of structural defects influence on the frequency and temperature dependences
of the ultrasound attenuation and dispersion.

\begin{acknowledgments}
This work was supported by the Ministry of Education and Science of Russia
through Grant No. 2.1.1/930.
\end{acknowledgments}


\begin{thebibliography}{99}
\bibitem{IkushimaF}
    A.~Ikushima and R.~Feigelson, J. Phys. Chem. Solids. \textbf{32}, 417 (1971).
\bibitem{Aliev}
    Kh.~K.~Aliev, I.~Kh.~Kamilov, and A.~M.~Omarov, Zh. \'{E}ksp. Teor. Fiz.
    \textbf{95}, 1896 (1989) [Sov. Phys. JETP \textbf{68}, 1096 (1989)].
\bibitem{LandauKh}
    L.~D.~Landau and I.~M.~Khalatnikov, Dokl. Akad. Nauk SSSR \textbf{96}, 496 (1954).
\bibitem{Pawlak}
    A.~Pawlak, Phys. Rev. B \textbf{44}, 5296 (1991).
\bibitem{Schwabl93}
    A.~M.~Schorgg and F.~Schwabl, Phys. Rev. B \textbf{49}, 11682 (1994).
\bibitem{Kamilov98}
    I.~K.~Kamilov and Kh.~K.~Aliev, Usp. Fiz. Nauk \textbf{168},
    953 (1998) [Phys. Usp. \textbf{41}, 865 (1998)].
\bibitem{Bhatt}
    R.~A.~Ferrell, B.~Mirhashem, and J.~K.~Bhattacharjee, Phys. Rev. B \textbf{35}, 4662 (1987).
\bibitem{Luthi}
    T.~J.~Moran and B.~L{\"u}thi, Phys. Rev. B \textbf{4}, 122 (1971).
\bibitem{Suzuki82}
    M.~Suzuki and T.~Komatsubara, J. Phys. C \textbf{15}, 4559 (1982).
\bibitem{Folk_Hol}
    R.~Folk, Yu.~Holovatch and T.~Yavors'kii, Phys.\ Usp. {\bf 46}, 169 (2003)
    [Uspekhi Fiz. Nauk {\bf 173}, 175 (2003)].
\bibitem{Harris74}
    A.~B.~Harris, J. Phys. C \textbf{7}, 1671 (1974).
\bibitem{PawlakFecher89}
    A.~Pawlak and B.~Fechner, Phys. Rev. B \textbf{40}, 9324 (1989).
\bibitem{Prudnikov_JETPL97}
V.~V.~Prudnikov, A.~V.~Ivanov, A.~A.~Fedorenko, JETP Lett. \textbf{66}, 835 (1997).
\bibitem{Prudnikov_JETP98}
V.~V.~Prudnikov, S.~V.~Belim, A.~V.~Ivanov, E.~V.~Osintsev, and
A.~A.~Fedorenko, Zh. \'{E}ksp. Teor. Fiz. \textbf{114}, 972 (1998)
[Sov. Phys. JETP \textbf{87}, 527 (1998)].
\bibitem{Prudnikov_JETP99}
V.~V.~Prudnikov, P.~V.~Prudnikov, A.~A.~Fedorenko, JETP \textbf{89}, 325 (1999).
\bibitem{PrudnikovPR00}
    V.~V.~Prudnikov, P.~V.~Prudnikov and A.~A.~Fedorenko, Phys.Rev. B \textbf{62}, 8777 (2000).
\bibitem{PrudnikovPR01}
    V.~V.~Prudnikov, P.~V.~Prudnikov and A.~A.~Fedorenko, Phys.Rev. B \textbf{63}, 184201 (2001).
\bibitem{Prudnikov_JETP02}
V.~V.~Prudnikov, P.~V.~Prudnikov, JETP \textbf{95}, 550 (2002).
\bibitem{PrudnikovCM}
    P.~V.~Prudnikov, V.~V.~Prudnikov, J. Phys.: Condens. Matter. \textbf{17}, L485 (2005).
\bibitem{Prudnikov_PhMM}
    P.~V.~Prudnikov, V.~V.~Prudnikov and E.~A. Nosikhin,
    Phys. Met. Metallogr. \textbf{104}, 221 (2007)
    [Fiz. Met. Metalloved. \textbf{104}, 235 (2007)].
\bibitem{Prudnikov_JETP}
    P.~V.~Prudnikov, V.~V.~Prudnikov, and E.~A.~Nosikhin, JETP
    \textbf{106}, 897 (2008) [Zh. \'{E}ksp. Teor. Fiz. \textbf{133}, 1027 (2008)].
\bibitem{WH}
A.~Weinrib and B.I.~Halperin, Phys.Rev. B \textbf{27}, 413 (1983).
\bibitem{Korzhenevskii}
A.L.~Korzhenevskii, A.A.~Luzhkov and W.~Schirmacher, Phys.Rev. B
\textbf{50}, 3661 (1994).
\bibitem{Dorogovtsev}
S.N.~Dorogovtsev, J.\ Phys.\ A {\bf 17}, L677 (1984).
\bibitem{Uzunov}
E.~Korutcheva and D.~Uzunov, Phys.\ Status\ Solidi\ (b) {\bf 126}, K19 (1984).
\bibitem{Korucheva}
E.~Korutcheva and F.~Javier de la Rubia, Phys.Rev. B \textbf{58}, 5153
(1998).
\bibitem{Prudnikov_PTP}
V.V.~Prudnikov, P.V.~Prudnikov, B.~Zheng, S.V.~Dorofeev, and V.Yu.~Kolesnikov,
Prog. Theor. Phys. {\bf 117}, 973 (2007).
\bibitem{Binder}
K.~Binder and J.D.~Reger, Adv.\ Phys. {\bf 41}, 547 (1992).
\bibitem{Blavats'ka}
V.~Blavats'ka, C.~von~Ferber and Yu.~Holovatch, Phys. Rev. E {\bf
64}, 041102 (2001).
\bibitem{Altarelli}
M.~Altarelli, M.D.~Nunez-Regueiro and M.~Papoular, Phys. Rev. Lett.
{\bf 74}, 3840 (1995).
\bibitem{He4}
J.~Yoon, D.~Sergatskov, J.~Ma, N.~Mulders, and M.~H.~W. Chan, Phys. Rev. Lett. {\bf 80}, 1461 (1998);
M.~Chan, N.~Mulders, and J.~Reppy, Phys. Today {\bf 49} (8), 30 (1996);
C.~V\'{a}squez~R., R.~Paredes~V., A.~Hasmy, and R.~Jullien,
Phys. Rev. Lett. {\bf 90}, 170602 (2003).
\bibitem{Larkin69}
    A.~I.~Larkin and S.~A.~Pikin, Zh. \'{E}ksp. Teor. Fiz. \textbf{56},
    1664 (1969) [Sov. Phys. JETP \textbf{29}, 891 (1969)].
\bibitem{Ymry74}
    Y.~Imry, Phys. Rev. Lett. \textbf{33}, 1304 (1974).
\bibitem{Izym}
    Yu.~A.~Izyumov and V.~N.~Syromyatnikov, Phase Transitions
    and Crystal Symmetry (Nauka, Moscow/Kluwer, Dordrecht, 1990).
\bibitem{MSI}
G.~Meissner, Ferroelectrics {\bf 24}, 27 (1980); F.~Schwabl and H.~Iro, Ferroelectrics {\bf 35}, 27 (1981).
\bibitem{IroSchwabl}
    H.~Iro, F.~Schwabl, Solid State Commun. \textbf{46}, 205 (1983).
\bibitem{Nelson76}
    D.~R.~Nelson,  Phys. Rev. B \textbf{14}, 1123 (1976).
\bibitem{Matching}
    R.~Folk, H.~Iro, F.~Schwabl, Z. Phys. B \textbf{27}, 169 (1977).
\bibitem{Kawasaki}
K.~Kawasaki, 1976, Phase Transitions and Critical Phenomena, (Academic London), Vol. 5a, p. 165;
K.~Kawasaki, 1977, Proceedings of Conference on Internal Friction and Ultrasonic Attenuation in Solids
(Tokyo University Press), p. 29.
\bibitem{Prudnikov01}
    V.~V.~Prudnikov and S.~V.~Belim, Fiz. Tverd. Tela (St.Petersburg) \textbf{43}, 1299 (2001) [Phys. Solid State \textbf{43}, 1353 (2001)].
\bibitem{Amit}
D.~Amit, Field Theory, The Renormalization Group, and Critical
Phenomena (McGraw-Hill, New York, 1978).
\bibitem{Zinn-Justin}
J.~Zinn-Justin, Quantum Field Theory and Critical Phenomena
(Clarendon Press, Oxford, 1996).
\bibitem{Sokolov}
S.~A.~Antonenko and A.~I.~Sokolov, Phys. Rev. B \textbf{49}, 15 901 (1994)
A.~I.~Sokolov, K.~B.~Varnashev, and A.~I.~Mudrov, Int. J. Mod. Phys. B \textbf{12}, 1365 (1998).
\bibitem{Baker}
G.~A.~Baker, B.~G.~Nickel, and D.~I.~Meiron, Phys. Rev. B \textbf{17}, 1365 (1978).
\bibitem{HH}
P.~C.~Hohenberg and B.~I.~Halperin, Rev. Mod. Phys. \textbf{49}, 435 (1977).
\bibitem{Rosov}
N.~Rosov, C.~Hohenemser, and M. Eibschutz, Phys. Rev. B \textbf{46}, 3452 (1992).
\end{thebibliography}
\end{document}